\documentclass[article,superscriptaddress,aps,twocolumn,pra,longbibliography]{revtex4-1}
\usepackage{dcolumn}
\usepackage{bm}
\usepackage{epsfig}
\usepackage{graphicx}
\usepackage{amssymb,amsmath,amsbsy,amsgen,amsfonts,amsthm,mathtools}    
\usepackage{mathrsfs}
\usepackage{latexsym}
\usepackage{array}
\usepackage{color}
\usepackage{amstext}
\allowdisplaybreaks[1]
\usepackage{txfonts}
\usepackage{pifont}
\usepackage{verbatim}
\usepackage{hyperref}

\usepackage{epstopdf} 
\usepackage{todonotes} 

\usepackage{soul}
\usepackage{ulem}

\newcommand{\abs}[1]{\left\vert #1\right\vert}
\newcommand{\bra}[1]{\left\langle{#1}\right\vert}
\newcommand{\ket}[1]{\left\vert{#1}\right\rangle}

\newcommand{\ave}[1]{\left\langle{#1}\right\rangle}

\DeclareSymbolFont{symbols}{OMS}{cmsy}{m}{n}

\begin{document}
\title{Experimental quantum polarimetry using heralded single photons}

\author{Seung-Jin Yoon}
\affiliation{Department of Physics, Hanyang University, Seoul, 04763, Korea}

\author{Joong-Sung Lee}
\affiliation{Department of Physics, Hanyang University, Seoul, 04763, Korea}

\author{Carsten Rockstuhl}
\affiliation{Institute of Theoretical Solid State Physics, Karlsruhe Institute of Technology, 76131 Karlsruhe, Germany}
\affiliation{Institute of Nanotechnology, Karlsruhe Institute of Technology, 76021 Karlsruhe, Germany}

\author{Changhyoup Lee}
\email{changhyoup.lee@gmail.com}
\affiliation{Institute of Theoretical Solid State Physics, Karlsruhe Institute of Technology, 76131 Karlsruhe, Germany}

\author{Kwang-Geol Lee}
\email{kglee@hanyang.ac.kr}
\affiliation{Department of Physics, Hanyang University, Seoul, 04763, Korea}

\date{\today}

\begin{abstract}
We perform experimental quantum polarimetry by using heralded single-photon to analyze the optical activity for linearly polarized light traversing a chiral medium. Three kinds of estimators are considered to estimate the concentration of sucrose solutions from measuring the rotation angle of the linear polarization of the output photons. Through repetition of independent and identical measurements performed for each individual scheme and different concentrations of sucrose solutions, we compare the estimation uncertainty among the three schemes. Results are also compared to classical benchmarks for which a coherent state of light is taken into account. The quantum enhancement in the estimation uncertainty is evaluated and the impact of experimental and technical imperfections is discussed. With our work, we lay out a route for future applications relying on quantum polarimetry. 
\end{abstract}


\maketitle

\section{Introduction}

\begin{figure*}[t]
\centering
\includegraphics[width=0.85\textwidth]{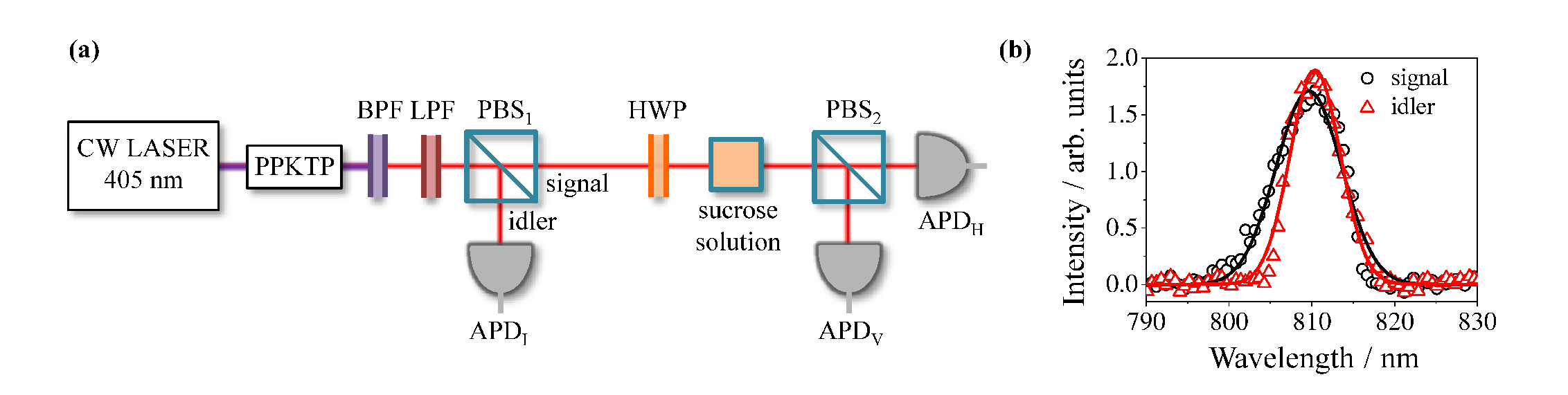}
\caption{
(a) Experimental scheme: A continuous-wave (CW) diode laser pumps a nonlinear crystal (a periodically poled potassium-titanyl-phosphate, PPKTP) to produce photon pairs. The paired photons are orthogonally polarized and separated by the first polarizing beam-splitter (PBS$_1$). The pump beam is removed from the optical system using a band pass filter (BPF) and a long pass filter (LPF). The detection of a vertically (V) polarized photon at an avalanche photodiode in the idler channel~($\text{APD}_\text{I}$) heralds the generation of a horizontally (H) polarized photons in the signal channel. The incident H-photons are rotated at a defined angle by a half-wave plate (HWP), and sent to a cuvette containing the sucrose solution under investigation. The polarization of the outgoing photons is analyzed via detection at~$\text{APD}_\text{H,V}$ placed in the output ports of the PBS$_2$. (b) The spectra of the photons generated by the spontaneous parametric downconversion process in the PPKTP. The central wavelengths of the signal and the idler photons are located at 809.6 nm and 810.4 nm, respectively.}
\label{fig:setup_spectrum}
\end{figure*}

Optical activity causes rotation of the linear polarization of light when it traverses a chiral material i.e., a material made from entities that lack mirror symmetry. Measurement of the optical activity has often been used in pharmacology, where it is important to study the chirality of drug molecules that determines their toxicity and efficacy, originated from the relative handedness of the enantiomers~\cite{Hutt1996, Zhao2017}. Chirality has also been explored in various fields, including chemistry, life science, physics, and material science, where polarimetric schemes provide rich information about molecular or nanophotonic chiral structures~\cite{Lough2002, Amabilino2009,  Barron2009, schaferling2017, hentschel2017, fernandez2019}.

A variety of detection schemes have been developed for measuring the optical activity, i.e., analyzing the rotation of the linear polarization of light upon propagation through a chiral medium~\cite{Carroll1963, Chou2006, Castiglioni2011}. This rotation of the polarization has its origin in the different propagation constants for the two counter-rotating circularly polarized plane waves, which are eigenmodes of the chiral medium and into which the linear polarization can be decomposed. The estimated direction of linear polarization would fluctuate due to the discrete nature of light even when the chirality is fixed and experimental noises are removed. These fluctuations determine the reliability of the measurement, often called the precision that reflects the data quality. In general, the precision increases with the intensity~$N$ (or the average photon number) of the light used. However, there exist situations that the intensity of an incident light impinging on an analyte is required to be limited when sample damage or any unwanted side effects occur in the high intensity regime, e.g., when the non-linear dependency is significant. Also, light itself can trigger unwanted chemical reactions in the molecular material to be detected~\cite{Neuman1999, Peterman2003, Taylor2015, Taylor2016}. The presence of such constraints has inspired the development of quantum metrological approaches over the last few decades. There, the aim is to improve the data quality while keeping the incident power in the low intensity regime~\cite{Wolfgramm2010, Roccia2018}, i.e., improving the precision for a fixed intensity.

Probing analytes with quantum states of light promises the detection with a much reduced noise below what is possible with classical states of light, thus allowing better performances in optical measurements~\cite{Giovannetti2004, Paris2009, Giovannetti2011}. Particularly, the phase-sensitive N00N states have been adopted to measure the chirality. For example, the Faraday rotation in a rubidium vapor cell has been measured with polarization N00N states, showing that the standard quantum limit can be beaten~\cite{Wolfgramm2013}. The dispersion of the optical rotation has also been analyzed by injecting polarization-entangled states with different and tunable wavelengths~\cite{Tischler2016}. Chemical processes that change the chirality of molecules have been monitored in real-time with polarization N00N states with a protocol in the context of quantum phase estimation~\cite{Cimini2019a, Cimini2019b}. In addition to schemes measuring the chirality,  N00N states have been employed to achieve a quantum enhancement in other sensing scenarios such as the measurement of the refractive index of blood proteins~\cite{Crespi2012}, in quantum lithography~\cite{D'Angelo2001}, and super-sensitive microscopy~\cite{Ono2013, Israel2014}. The superior optical detection shown with the N00N states originates from the quantum entanglement of photons, leading to a higher signal-to-noise ratio~($\sqrt{N}$ times bigger) when compared to measurements that rely on classical sources such as coherent states of light. The quantum advantage enabled by N00N states is known to be vulnerable to photon loss or decoherence~\cite{Dowling2008, chen2018}, but one can use optimally engineered definite photon-number states that always outperform the standard quantum limit for a given loss~\cite{Dorner2009, Kacprowicz2010, lee2016}.

In the context of polarimetry the chirality of a molecular solution is measured by measuring the transmittance~$T$ of a linearly polarized incident light passing through an analyte resolved in its co- or cross-polarized components. For definiteness we speak in the following of horizontal and vertical polarization. The presence of the chiral medium rotates weakly the polarization, which can be probed from the change of the intensity of the incidence polarization or from the difference between horizontally and vertically polarized light. From these measured intensities we can conclude on the concentration of the molecules if the chirality itself is known. It is known that the photon number state~$\ket{N}$ is the optimal state that minimizes the noise in the transmission measurement~\cite{monras2007,alipour2014}. That is because the photon number state has no uncertainty in its intensity, offering the most precise detection when monitoring the change of the intensity~\cite{Meda2017, adesso2009}. Unlike N00N states, the photon number state always offers a quantum gain in noise reduction at any loss level~$\gamma~(=1-\eta)$, exhibiting a detection fluctuation~$\Delta_\text{q} T=\sqrt{T(1-\eta T)/N\eta}$ compared to~$\Delta_\text{c} T=\sqrt{T/N\eta}$ that is obtainable when using coherent states with an average photon number of~$N$. Therefore, the photon number state is not just optimal in reducing the noise or uncertainty, but also useful against the photon loss in realistic transmission measurements. As generation of Fock states with an arbitrary photon number~$N$ is not yet available with current technology, one can use instead~$N$~single photons relying on the equivalence to the case using an~$N$ photon number state in transmission measurement. Several studies have  been carried out, such as absorption spectroscopy to analyze the organic dye molecule dibenzanthanthrene~\cite{Rezus2012} or haemoglobin~\cite{Whittaker2017}, surface plasmon resonance sensing to measure the change of the refractive index of an analyte under investigation with such Fock states~\cite{Lee2018}.

In this work, we study theoretically and experimentally how much quantum enhancement can be obtained in measuring the optical activity of chiral molecular materials with single-photons as an input state. 
We use a heralding scheme to generate the single photon state that is linearly polarized and that illuminates sucrose solutions with different concentrations. To estimate the concentration of sucrose solutions, we consider three typical polarimetric schemes: measurement of (i) the intensity of the horizontally polarized outgoing photons, (ii) the intensity difference between horizontally and vertically polarized outgoing photons, and (iii) the intensity difference-to-sum ratio between the two linearly polarized outgoing photons. 
We analyze and compare the performances of these three schemes that use single photon states among each others but also to the optimal precision achievable with a classical input state. 
We find that type-(iii) leads to the minimal estimation uncertainty as compared to the other two types. A quantum enhancement, i.e., an improvement of the estimation precision when compared to the measurement with a classical state of light is experimentally observed in both type-(i) and type-(ii) measurements. In general, our study proves that the quantum enhancement is always obtained at any~$\eta$ and~$T$. The effect of experimental imperfections is also discussed, e.g., the extinction ratio of the optical components used in the experiment.

\section{Experimental scheme}\label{sec:phase_estimation}

The experimental scheme used in this work is shown in Fig.~\ref{fig:setup_spectrum}(a). A continuous wave (CW) diode laser with a wavelength of 405~nm, filtered out at a later stage by a band pass filter (BPF) and a long pass filter (LPF), pumps a nonlinear crystal (a periodically poled potassium-titanyl-phosphate, PPKTP) to generate paired photons at the emission wavelength of 809.6 nm and 810.4 nm with a full width at half maximum of 9.4 nm and 7.2 nm, respectively [see Fig.~\ref{fig:setup_spectrum}(b)]. The two orthogonally polarized photons are spatially separated by a polarizing beam-splitter (PBS). One of the two photons, the vertically polarized idler photon, is directly sent to an avalanche photodiode single-photon detector (APD, SPCM-AQR-15, PerkinElmer). Quantum correlation between photon pairs enables the click events from the idler detector APD\textsubscript{I} to herald the horizontally polarized signal photon that is fed into the polarimetry setup under investigation. Rotation of the linear polarization occurs when the signal photon passes through a sucrose solution. The magnitude of this rotation depends on the concentration of the sucrose solution, eventually determining the extent to which the polarization of the outgoing photons rotates. To make the scheme operating with heralded single photons in detection, we adopt coincident measurement between  APD\textsubscript{I} and  APD\textsubscript{H,V}, i.e., the results of detection at   APD\textsubscript{H,V} are recorded only when the single photons are detected at  APD\textsubscript{I}. Throughout this paper, let~$N_{\text{H,V}}$ be the coincidence counts between  APD\textsubscript{I} and  APD\textsubscript{H,V}.

The heralded single photon scheme is valid in the low gain parameter regime, where the state generated via SPDC can be written as~$\ket{\text{SPDC}}\approx\ket{00}+\epsilon\ket{11}\text{with} \abs{\epsilon}\ll1$~\cite{Whittaker2017}. The dead time of the APDs used in our experiment is~$\sim$~60 ns and the time window of the field programmable gate array (FPGA), used for time-tracking analysis, is 25 ns. Considering these time-scales, we set the count rate of the idler photons to be about~$8\times10^5$~cps by tuning the intensity of the pump laser. The sample size is set to~$\nu =10^5$ corresponding to the number of the detection events in the idler port (APD\textsubscript{I}), which heralds the twin photon in the signal port. We repeat the independent and identical measurement~$\mu =500$ times, assumed to be large enough to extract reliable statistical features of interest. Such sampling is applied to each polarization angle of incidence and each concentration of the sucrose solution. 

The signal H-photons are rotated through a half wave plate (HWP), controlling the input polarization of~$\theta_\text{in}$ in a range from~$-100^\circ$ to~$100^\circ$ in steps of 10$^\circ$. The input photons being linearly polarized at~$\theta_\text{in}$ pass through sucrose solutions with the concentrations~$C$ that vary from 0.1~g~ml$^{-1}$ to 0.6~g~ml$^{-1}$ in steps of 0.1~g~ml$^{-1}$. The mirror asymmetry of sucrose molecules induces a rotation of the linear polarization by~$\alpha$ that depends in its magnitude on the concentration~$C$, resulting in outgoing photons with a linear polarization at~$\theta_\text{out}=\theta_\text{in}+\alpha$. The latter is analyzed through decomposition into the H- and V-polarizations, being realized by the PBS. 
When either~$N_{\text{H}}$ or~$N_{\text{V}}$ is only measured, our scheme is equivalent to the conventional polarimetric scheme, where a linear polarizer is inserted and rotated before a detector while keeping an incident polarization the same.

The rotation angle~$\alpha$ of the linear incident polarization as an effect of the optical activity is proportional to the solution concentration~$C$ and the propagation length~$l$ through the material, so the angle~$\alpha$  can be written as
\begin{align}
\alpha(\lambda)=[\alpha(\lambda)]\times l \times C, \label{eq:opt_act}
\end{align}
where~$[\alpha(\lambda)]$ is the specific rotation of the sample material. The wavelength dependence in the specific rotation can be modelled by Drude's expression, written as
\begin{align}
[\alpha(\lambda)]=\sum_{j}{\frac{A_j}{\lambda^2-\lambda_j^2}}, \label{eq:spec_rot}
\end{align}
where the summation is over multiple excitation transitions with~$\lambda_j$ and~$A_j$ being the resonance wavelength and the rotation amplitude for the~$j^\text{th}$ transition, respectively. For the case of sucrose, the dispersion of the optical activity can be characterized by a single transition (i.e.,~$j=1$), for which~$A_1=2.1648\times10^7 ~\text{deg}~\text{nm}^2~\text{dm}^{-1}~\text{g}^{-1}~\text{ml}$ and~$\lambda_1=146~\text{nm}$~\cite{Tischler2016, Lowry1924}. We thus aim to estimate the concentration~$C$ from the measurement of the angle~$\alpha(\lambda)$ for~$l=0.1~\text{dm}$ (the length of the cuvette used in our experiment) and~$[\alpha(\lambda)]$ given above.

\begin{figure}[t]
\centering
\includegraphics[width=0.44\textwidth]{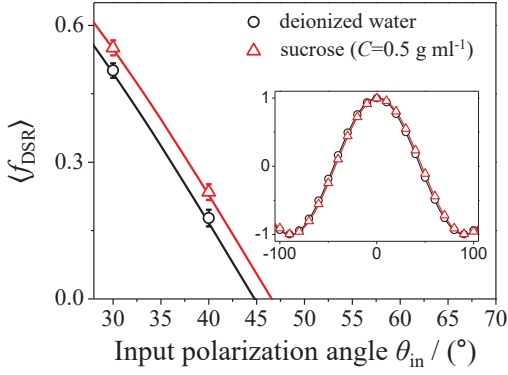}
\caption{
Open symbols show the measured~$\ave{f_\text{DSR}}$ of Eq.~\eqref{eq:f_DSR} while varying the input polarization angle~$\theta_\text{in}$ for deionized water (black circles) and for sucrose solution with~$C=0.5$~g~ml$^{-1}$ (red triangles). 
The inset shows the measured values in a full range of~$\theta_\text{in}$ from~$-100^\circ$ to~$100^\circ$ in steps of~$10^\circ$.
The horizontal displacement between the two sets of data exhibits the optical activity induced by the chiral material, enabling estimation of the rotation angle~$\alpha$. Solid curves are the fitted values using Eq.~\eqref{eq:pre_DSR_q}. 
The error bars are the standard deviation in the histogram over~$\mu$ times of repetition at each value of~$\theta_\text{in}$.
}
\label{fig:scanning}
\end{figure}

In this work, we implement three kinds of estimator yielding the associated values~$f$'s, which are often used in classical polarimetry~\cite{Carroll1963, Chou2006, Castiglioni2011}: (i) the number~$N_{\text{H}}$ of the horizontally polarized photons transmitted through the sample, but normalized by~$\nu$. (ii) the difference~$(N_{\text H}-N_{\text V})$ between the horizontally and the vertically polarized photons transmitted through the sample, normalized by~$\nu$. (iii) the difference~$(N_{\text H}-N_{\text V})$ normalized by the sum~$(N_{\text H}+N_{\text V})$, i.e., the difference-to-sum ratio (DSR) which has recently been employed in imaging of non-uniform refractive profiles~\cite{ortolano2020}. 
The expectation values of the above three estimators for~$\nu$ single photons of incidence can be written respectively~as
\begin{align}
\text{(i)}\text{  }\left\langle{f_\text{single}}\right\rangle&=\frac{\ave{N_\text{H}}}{\nu}=\eta_\text{H}~T_{\theta_\text{out}},\label{eq:f_single}\\
\text{(ii)  }\text{ }\text{ }\left\langle{f_\text{diff}}\right\rangle&=\frac{\ave{N_\text{H}-N_\text{V}}}{\nu}=\eta_\text{H}~T_{\theta_\text{out}}-\eta_\text{V}~R_{\theta_\text{out}},\label{eq:f_diff}\\
\text{(iii)}\text{ }\left\langle{f_\text{DSR}}\right\rangle&=\ave{\frac{N_\text{H}-N_\text{V}}{N_\text{H}+N_\text{V}}}
\simeq\frac{\eta_\text{H}~T_{\theta_\text{out}}-\eta_\text{V}~R_{\theta_\text{out}}}{\eta_\text{H}~T_{\theta_\text{out}}+\eta_\text{V}~R_{\theta_\text{out}}},\label{eq:f_DSR}
\end{align}
where~$\ave{..}$ denotes an average with respect to the output state being measured and~$\eta_\text{H,V}$ denote the efficiencies of transmission from the PPKTP to the detectors~$\text{APD}_\text{H,V}$ including the detection efficiencies. 
Here,~$T_{\theta_\text{out}}=\text{cos}^2\theta_\text{out}$ and~$R_{\theta_\text{out}}=\text{sin}^2\theta_\text{out}$ denote the transmittance and the reflectance for the outgoing single photon with a polarization of~$\theta_\text{out}=\theta_\text{in}+\alpha$ to be transmitted through and reflected from the PBS$_2$ in Fig.~\ref{fig:setup_spectrum}, respectively.
Particularly Eq.~\eqref{eq:f_DSR} asymptotically holds for a large sample size~$\nu$, according to Jensen's inequality~\cite{Jensen1906}, which is the case in our experiment. In an ideal case where~$\eta_\text{H}=\eta_\text{V}, N_\text{H}+N_\text{V}=\nu$, so that Eq.~\eqref{eq:f_DSR} becomes Eq.~\eqref{eq:f_diff}. The values~$\eta_\text{H,V}\approx0.25$ are explicitly measured by the coincidence count~$N_\text{H,V}$ with scanning over~$\theta_\text{in}$ when pure water is in the cuvette, i.e.,~$\alpha=0$ (see black curves and circles in Fig.~\ref{fig:scanning}), which exploits the correlated features of the SPDC photons as used in the Klyshko method~\cite{Klyshko1980, Burnham1970, Rarity1987}.

In what follows, Eqs.~\eqref{eq:f_single}-\eqref{eq:f_DSR} are used to estimate the angle of rotation~$\alpha$ from the measurement of the output polarization~$\theta_\text{out}=\theta_\text{in}+\alpha$ for a given input polarization~$\theta_\text{in}$. The estimation of~$\alpha$ subsequently leads to the estimation of the concentration~$C$ of sucrose solutions through Eq.~\eqref{eq:opt_act}. The impact of a finite concentration of the sucrose solution can be seen as a shift of the curve (for example, see red curves and triangles in Fig.~\ref{fig:scanning} for the measurement of the sucrose solution with a concentration of 0.5 g~ml$^{-1}$). Assuming the values of~$[\alpha]$,~$l$,~$\eta_\text{H,V}$, and~$\theta_\text{in}$ to be accurately known beforehand, one can determine the estimation uncertainty~$\Delta C$ directly by the estimation uncertainty of the output polarization~$\Delta \theta_\text{out}$ extracted from the measurement of the outcome~$f$'s.

\section{Results and Discussion}

The~$\Delta\theta_\text{out}$ is experimentally obtained by the standard deviation of the estimated values~$\theta_\text{out}$ from~$\mu$ measurements with a sample size of~$\nu$, i.e.,~$\Delta\theta_\text{out}^\text{exp}=[\sum_{j}\theta_\text{out}^2/\mu-(\sum_{j}\theta_\text{out}/\mu)^2]^{1/2}$. The experimentally measured value $\Delta\theta_\text{out}^\text{exp}$ is compared to the theoretical prediction~$\Delta\theta_\text{out}$, which can be calculated by considering the linear error propagation from the variance of the outcomes~$\Delta f=\langle(\Delta f)^2\rangle^{1/2}$~\cite{Braunstein1994}, written as
\begin{align}
\Delta\theta_\text{out}=\frac{\Delta f}{\abs{\frac{\partial\ave{f}}{\partial\theta_\text{out}}}}. \label{eq:LEPM}
\end{align}
The variance~$\langle(\Delta f)^2\rangle$ can be calculated as 
\begin{align}
\langle(\Delta f)^2\rangle=\sum_{j,k \in \{\text{H},\text{V}\}}\frac{\partial\ave{f}}{\partial N_j}\frac{\partial\ave{f}}{\partial N_k} \text{Cov}(N_j,N_k),\label{eq:noise_expansion}
\end{align}
where~$\text{Cov}(N_\text{j},N_\text{k})$ is the covariance between the photon counts~$N_\text{j}$ and~$N_\text{k}$. Eq.~\eqref{eq:noise_expansion} is an exact function for~$f_\text{single}$ and~$f_\text{diff}$, but is an approximate function for~$f_\text{DSR}$ since Eq.~\eqref{eq:f_DSR} is a non-linear differential function of~$N_\text{H,V}$~\cite{Abayzeed2018}. In our experiment,~$\nu$ heralded single photons are used, and the theoretical expression of~$\Delta \theta_\text{out}$ of Eq.~\eqref{eq:LEPM} can be written respectively as
\begin{align}
\Delta\theta_\text{out}^\text{single(q)}&=\frac{1}{\sqrt{\nu}}\sqrt{\frac{1-\eta_\text{H}T_{\theta_\text{out}}}{4\eta_\text{H}(1-T_{\theta_\text{out}})}},\label{eq:pre_sing_q}\\
\Delta\theta_\text{out}^\text{diff(q)}&=\frac{1}{\sqrt{\nu}}\sqrt{\frac{(\eta_\text{H}T_{\theta_\text{out}}+\eta_\text{V}R_{\theta_\text{out}})-(\eta_\text{H}T_{\theta_\text{out}}-\eta_\text{V}R_{\theta_\text{out}})^{2}}{4(\eta_\text{H}+\eta_\text{V})^2T_{\theta_\text{out}}R_{\theta_\text{out}}}},\label{eq:pre_diff_q}\\
\Delta\theta_\text{out}^\text{DSR(q)}&=\frac{1}{\sqrt{\nu}}\sqrt{\frac{\eta_\text{H}T_{\theta_\text{out}}+\eta_\text{V}R_{\theta_\text{out}}}{4\eta_\text{H}\eta_\text{V}}}.\label{eq:pre_DSR_q}
\end{align}
\begin{figure}[]
\centering
\includegraphics[width=0.45\textwidth]{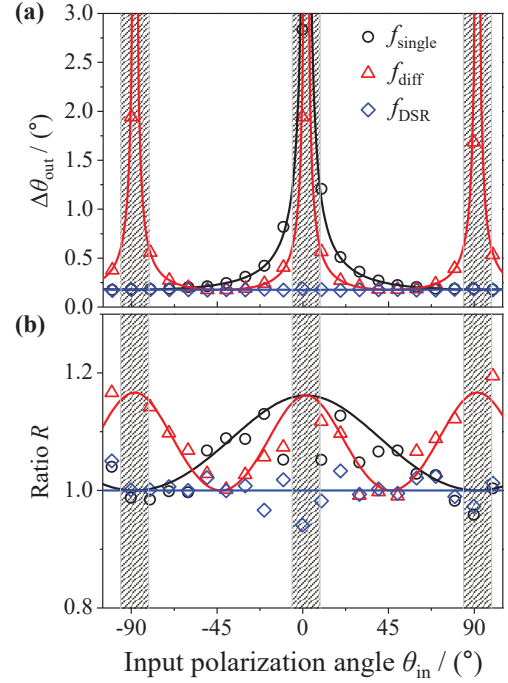}
\caption{
(a) Experimentally measured estimation uncertainties~(open symbols) of the output polarization angle~$\Delta\theta_\text{out}$ with the three estimation schemes introduced in the main text, i.e.,~$\Delta\theta_\text{out}^\text{DSR(q)}$,~$\Delta\theta_\text{out}^\text{single(q)}
$, and~$\Delta\theta_\text{out}^\text{diff(q)}$. They are in good agreement with theoretical predictions~(solid curves) using Eqs.~\eqref{eq:pre_sing_q}-\eqref{eq:pre_DSR_q}. The three schemes achieve almost the same minimal uncertainties since~$\eta_\text{H}\approx\eta_\text{V}$ holds in our experiment, as expected from theoretical analysis. 
(b) Relative quantum enhancement~$R$ with respect to the estimation uncertainties that would be obtained when probing the optical activity with a coherent state of light. Solid curves represent theoretical ratios~$R$'s calculated by using Eqs.~\eqref{eq:pre_sing_q}-\eqref{eq:pre_DSR_q} and Eqs.~\eqref{eq:pre_sing_c}-\eqref{eq:pre_DSR_c}.
Importantly, the shaded regions are excluded, where the estimation is unphysical or malfunctioning due to a non-ideal extinction ratio of the PBS used in the experiment.  
}\label{fig:preci_enhance}
\end{figure}

The uncertainties~$\Delta\theta_\text{out}^\text{exp}$ are experimentally measured by using heralded single photons in the three schemes with varying~$\theta_\text{in}$ for the sucrose solution with a concentration of 0.5 g~ml$^{-1}$ [see open symbols in Fig.~\ref{fig:preci_enhance}(a)]. The experimentally measured uncertainties are compared with those calculated using Eqs.~\eqref{eq:pre_sing_q}-\eqref{eq:pre_DSR_q} [see solid curves in Fig.~\ref{fig:preci_enhance}(a)], providing the analyses for the observed notable behaviors as a function of~$\theta_\text{in}$.

The two uncertainties~$\Delta\theta_\text{out}^\text{single(q)}$ and~$\Delta\theta_\text{out}^\text{diff(q)}$ diverge as~$T_{\theta_\text{out}}\rightarrow 1$ (unless~$\eta_\text{H}=1$) and ~$T_{\theta_\text{out}}\rightarrow 0 ~\text{or}~1$ (unless~$\eta_\text{H}=\eta_\text{V}=1$), but decrease as~$T_{\theta_\text{out}}\rightarrow 0$ and~$T_{\theta_\text{out}}\rightarrow \eta_\text{V}(1-\eta_\text{V})/[\eta_\text{V}(1-\eta_\text{V})+\sqrt{\eta_\text{H}(1-\eta_\text{H})\eta_\text{V}(1-\eta_\text{V})}]$, respectively. At the latter limits, they have the minima:~$\text{min}[\Delta\theta_\text{out}^\text{single(q)}]=1/(\sqrt{\nu}\sqrt{4\eta_\text{H}})$ and~$\text{min}[\Delta\theta_\text{out}^\text{diff(q)}]=[\eta_\text{H}(1+\eta_\text{V})+\eta_\text{V}(1+\eta_\text{H})+2\sqrt{\eta_\text{H}(1-\eta_\text{H})\eta_\text{V}(1-\eta_\text{V})}]^{1/2}/(2\sqrt{\nu}(\eta_\text{H}+\eta_\text{V}))$. On the other hand, the uncertainty~$\Delta\theta_\text{out}^\text{DSR(q)}$ has the minimum~$\text{min}[\Delta\theta_\text{out}^\text{DSR(q)}]=1/(\sqrt{\nu}\sqrt{4\eta_\text{H}})$ at~$T_{\theta_\text{out}}=0$ and the maximum~$\text{max}[\Delta\theta_\text{out}^\text{DSR(q)}]=1/(\sqrt{\nu}\sqrt{4\eta_\text{V}})$ at~$T_{\theta_\text{out}}=1$ when~$\eta_\text{H}>\eta_\text{V}$. The minima and maxima are reversed when~$\eta_\text{V}>\eta_\text{H}$. When~$\eta_\text{H}=\eta_\text{V}\equiv\eta$, the uncertainty~$\Delta\theta_\text{out}^\text{DSR(q)}$ takes a constant value~$1/(\sqrt{\nu}\sqrt{4\eta})$ regardless of~$\theta_\text{in}$. Also note that
\begin{align}
\text{min}[\Delta\theta_\text{out}^\text{DSR(q)}]=\text{min}[\Delta\theta_\text{out}^\text{single(q)}]< \text{min}[\Delta\theta_\text{out}^\text{diff(q)}]
\end{align}
when~$\eta_\text{H}~>~\eta_\text{V}$, whereas 
\begin{align}
\text{min}[\Delta\theta_\text{out}^\text{DSR(q)}]<\text{min}[\Delta\theta_\text{out}^\text{single(q)}]< \text{min}[\Delta\theta_\text{out}^\text{diff(q)}]
\end{align}
when~$\eta_\text{V}>\eta_\text{H}$. For the equal efficiencies~$\eta_\text{V}=\eta_\text{H}$,~$\text{min}[\Delta\theta_\text{out}^\text{DSR(q)}]=\text{min}[\Delta\theta_\text{out}^\text{single(q)}]=\text{min}[\Delta\theta_\text{out}^\text{diff(q)}]$. All implies that the DSR scheme using Eq.~\eqref{eq:f_DSR} generally provides the optimal sensing scheme with the minimal uncertainty among the three estimators in any condition. The aforementioned theoretically expected behaviors are manifested in the experimentally measured values~$\Delta\theta_\text{out}^\text{exp}$ and shown in Fig.~\ref{fig:preci_enhance}(a)].

The classical benchmarks obtained with a coherent state with an average photon number of~$\langle N\rangle =1$ for the individual schemes are given by 
\begin{align}
\Delta\theta_\text{out}^\text{single(c)}&=\frac{1}{\sqrt{\nu}}\sqrt{\frac{1}{4\eta_\text{H}(1-T_{\theta_\text{out}})}},\label{eq:pre_sing_c}\\
\Delta\theta_\text{out}^\text{diff(c)}&=\frac{1}{\sqrt{\nu}}\sqrt{\frac{\eta_\text{H}T_{\theta_\text{out}}+\eta_\text{V}R_{\theta_\text{out}}}{4(\eta_\text{H}+\eta_\text{V})^2 T_{\theta_\text{out}}R_{\theta_\text{out}}}},\label{eq:pre_diff_c}\\
\Delta\theta_\text{out}^\text{DSR(c)}&=\frac{1}{\sqrt{\nu}}\sqrt{\frac{\eta_\text{H}T_{\theta_\text{out}}+\eta_\text{V}R_{\theta_\text{out}}}{4\eta_\text{H}\eta_\text{V}}},\label{eq:pre_DSR_c}
\end{align}
respectively. The relative quantum enhancement~$R$ can be quantified by the ratio of~$\Delta\theta_\text{out}^\text{(c)}$ to~$\Delta\theta_\text{out}^\text{(q)}$, written as
\begin{align}
R=\frac{\Delta\theta_\text{out}^\text{(c)}}{\Delta\theta_\text{out}^\text{(q)}}, \label{eq:q_enhance}
\end{align}
Therefore, a value of~$R$ greater than unity exhibits a quantum enhancement. For the three schemes considered above, one can show that~$R^\text{single}=(1-\eta_\text{H}T_{\theta_\text{out}})^{-1/2}$,~$R^\text{diff}=[1-(\eta_\text{H}T_{\theta_\text{out}}-\eta_\text{V}R_{\theta_\text{out}})/(\eta_\text{H}T_{\theta_\text{out}}+\eta_\text{V}R_{\theta_\text{out}})]^{-1/2}$, and~$R^\text{DSR}=1$. Interestingly~$R$ is always equal to or greater than unity for all schemes considered in our experiment. These behaviors of the ratio~$R$ are presented in Fig.~\ref{fig:preci_enhance}(b), given by the experimentally measured noise~$\Delta\theta_\text{out}^{(q)}$ and the theoretical value~$\Delta\theta_\text{out}^{(c)}$ that would be obtainable via the optimal classical polarimetry.

The estimators of Eqs.~\eqref{eq:f_single}-\eqref{eq:f_DSR} are unbiased under ideal conditions. One may wonder if the above estimation uncertainties evaluated using Eq.~\eqref{eq:LEPM} reach the lower bound given by Cram{\'e}r-Rao (CR) inequality being satisfied for any unbiased estimator~\cite{durkin2007, pezze2018}. The CR inequality can be written as~$\Delta \theta_\text{out} \ge 1/\sqrt{\nu F (\theta_\text{out})}$ with~$F(\theta_\text{out})$ being the Fisher information (FI) for a parameter~$\theta_\text{out}$. It depends on the measurement setting that is performed~\cite{Cramer1999}. From the expressions of FIs (see Appendix~\ref{apped:FI}) for both classical and quantum schemes considered in this work, one can clearly see that the uncertainties evaluated using Eq.~\eqref{eq:LEPM} are the same as the inverse of the squared FIs regardless of~$T_{\theta_\text{out}}$,~$R_{\theta_\text{out}}$, and~$\eta_\text{H,V}$ when estimating the parameter with the single-mode measurement scheme (i.e.,~$f_\text{single}$). However, the uncertainties of Eq.~\eqref{eq:LEPM} with the two-mode measurement schemes (i.e.,~$f_\text{diff}$ and~$f_\text{DSR}$) are equal to or larger than the inverse of the squared FIs. It implies that a better estimator than using~$f_\text{diff}$ and~$f_\text{DSR}$ exists when the photon-number-counting measurement is performed at the two modes.

When the measurement setting is optimized, the CR inequality can be further developed to be the one called the quantum CR inequality, where the maximized FI is called the quantum FI (QFI)~\cite{Braunstein1994, Braunstein1996}. For the output state decomposed into~$\hat{\rho}_{\theta_\text{out}}=\sum_j p_j \ket{\psi_j}\bra{\psi_j}$ with~$\langle \psi_j \vert \psi_k\rangle =\delta_{j,k}$, it is known that the FI is the same as the QFI if~$\partial_{\theta_\text{out}} \bra{\psi_j}=0~\forall j$. Interestingly, the latter is the case for both classical and quantum schemes considered in this work. It indicates that the photon-number-counting measurement is the optimal measurement scheme for both classical and quantum approaches. This may not be true for other kinds of probe states, i.e., an individual optimal measurement scheme needs to be found and performed for maximization of quantum enhancement when other types of quantum states are illuminated into the sample.

\begin{figure}[t]
\centering
\includegraphics[width=0.5\textwidth]{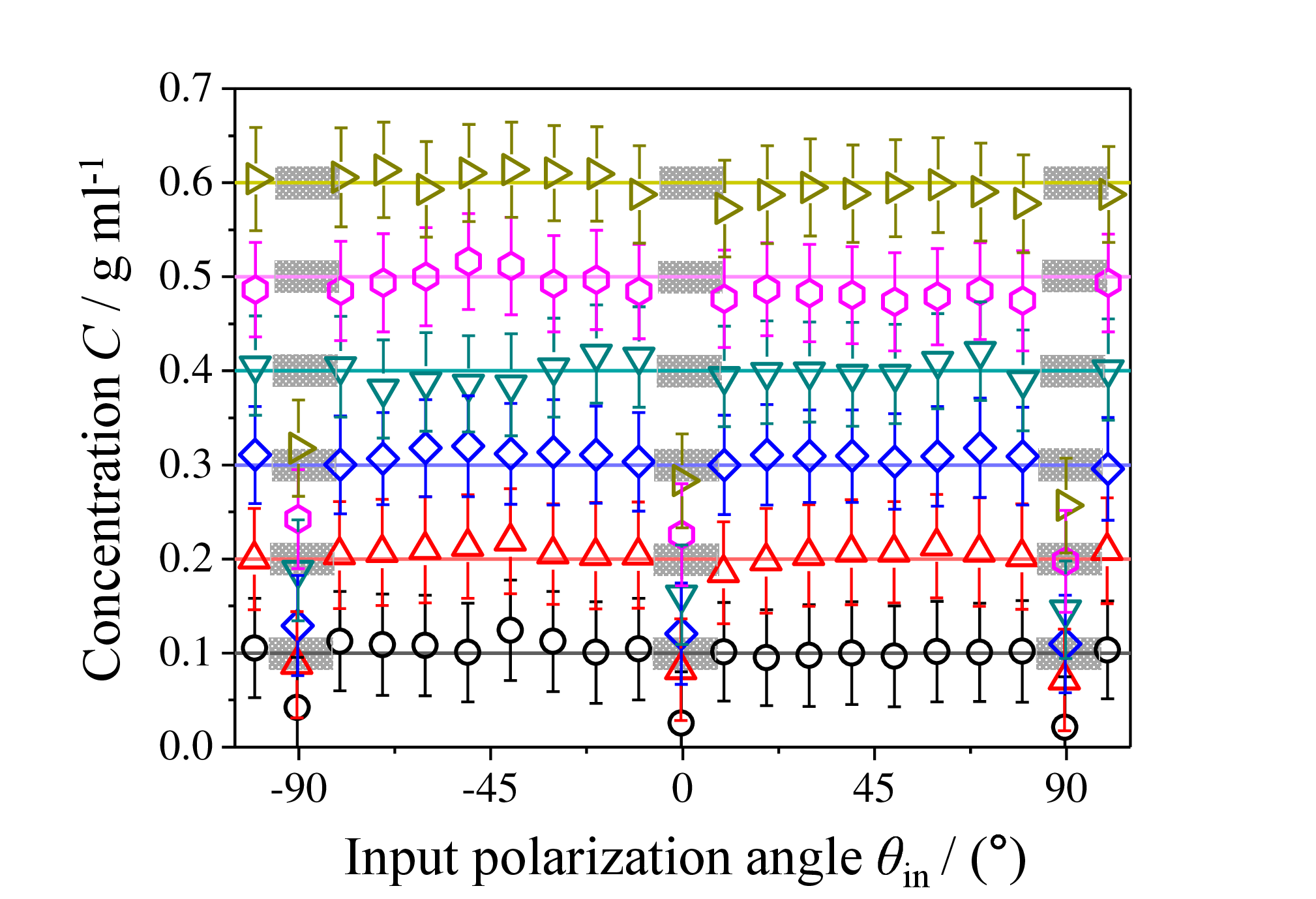}
\caption{
Experimentally estimated concentrations of sucrose solutions with six different concentrations are obtained by using the estimator of Eq.~\eqref{eq:f_DSR}. We exclude the shaded regions as in Fig.~\ref{fig:preci_enhance}, where the estimated concentrations are misled due to a non-ideal extinction ratio (see Appendix~\ref{MCsim} for verification made via the Monte-Carlo simulation). 
The error bar at individual points represents the standard deviation of the histogram obtained over~$\mu$ times measurements. 
}
\label{fig:concent}
\end{figure}

Actual estimation of the concentration of sucrose solutions can be made via the relation~$\theta_\text{out}=\langle \theta_\text{in}\rangle+\alpha$ and Eq.~\eqref{eq:opt_act}. Figure~\ref{fig:concent} shows the results of the scheme using the DSR estimator of Eq.~\eqref{eq:f_DSR} for sucrose solutions with different concentrations. The average estimated values~$C$ of the concentration are shown together with the estimation uncertainties~$\Delta C$. The latter reflects the type A uncertainties caused by probabilistic photon number counting. Further evaluation regarding uncertainties of other factors can be made, but we leave rigorous investigation of those minor contributions as a future work while providing the type B uncertainty budget in Appendix~\ref{uncer_bud}~(see Table~\ref{tab:uncer_bud_table}). The shaded areas in Fig.~\ref{fig:concent} represent fallible regions which would lead to erroneous estimation, so we exclude those regions from estimation.  


It is worth discussing two experimental imperfections that led to the exclusion of the fallible regions in Figs.~\ref{fig:preci_enhance} and \ref{fig:concent}. First, the polarization control by experimental components may not be ideal, e.g., an incident H-photon to PBS can exit through the V-photon output port. Its rate can be quantified by the polarization extinction (PE) ratio~$r_\text{PE}$, which is measured to be about 1000/1 in our experiment. This means that even when~$\theta_\text{out}=0$, a few photons are found in~$\text{APD}_\text{V}$ although no photon is supposed to be detected there in the ideal case. Such an imperfection becomes more significant as~$T_{\theta_\text{out}}\rightarrow 0\text{ or }1$, around which the estimation of the concentration~$C$ is likely to be more inaccurate. The effect of the PE is investigated via Monte-Carlo simulation~(see the details in Appendix~\ref{MCsim}). We also identify that the imperfect functioning of experimental components can contribute to type A uncertainty.
Second, particularly when the estimation is made with Eq.~\eqref{eq:f_single} at each incidence angle~$\theta_\text{in}$, the experimentally measured efficiency~$\eta_\text{H}$ is inserted into Eq.~\eqref{eq:f_single}. This works reasonably well except for the cases when more than~$\eta_\text{H}\nu$ photons are detected in the measurement counting the photon number~$N_\text{H}$. For example, consider the extreme yet possible case that all~$\nu$ photons are detected in~$N_\text{H}$ even when~$\eta_\text{H}<1$. Note that photon loss occurs probabilistically with a probability~$\eta_\text{H}$. In this case,~$N_\text{H}/\eta_\text{H}\nu$ can be greater than unity and thus~$T_{\theta_\text{out}}>1$, leading to unphysical estimation. This arises because the value~$N_\text{H}/\nu$ is divided by~$\eta_\text{H}$, implying that the probabilistic nature associated with the efficiency~$\eta_\text{H}$ is regarded as a deterministic process that supposes only a definite number~$\eta_\text{H}\nu$ of photons passes through an analyte, which is of course wrong and misses the probabilistic nature. Such an extreme case often take places when~$T_{\theta_\text{out}}\rightarrow 1$ and similarly when~$T_{\theta_\text{out}}\rightarrow 0$. The same concern also applies to the estimation with Eq.~\eqref{eq:f_diff}. Therefore, we exclude the range of~$\theta_\text{out}$ such that~($\eta_\text{H,V}\nu-\ave{N_\text{H,V}(\theta_\text{out})}<3\ave{\Delta N_\text{H,V}(\theta_\text{out})}$) for all the estimation schemes considered in this work in order to avoid inaccurate or unphysical estimation. The corresponding regimes are represented by shaded areas in Figs.~\ref{fig:preci_enhance} and \ref{fig:concent}.

The first issue could be alleviated by using optical components with as the higher PE ratio~$r_\text{PE}$ as possible or at least a much better ratio on a logarithmic scale. The second issue, on the other hand, requires a more sophisticated modification to the estimation schemes. For example, one could perform simultaneous estimation of the concentration~$C$ and the efficiencies~$\eta_\text{H,V}$ from the measurement with scanning through the incidence angles. This can typically be done by fitting Eqs.~\eqref{eq:f_single}-\eqref{eq:f_DSR} to the measured data, which we leave for future study.

\section{Conclusion}
We have experimentally investigated quantum polarimetric schemes using single-photon inputs to analyze the concentration of sucrose solutions. The horizontally and vertically polarized outgoing photons have been counted for each polarization of incidence and each concentration of sucrose solution in the experiment, leading to the determination of the optical activity. It has been shown that the minimal estimation uncertainty is generally achieved when the concentration is analyzed with the normalized difference between H- and V-photons, i.e., Eq.~\eqref{eq:f_DSR}. We have discussed the quantum gain in the detection of the optical activity with respect to the three typical polarimetric schemes and the effect of the experimental and technical imperfections is identified.

When probing the optical activity with more intense light than a single photon state and when it is, in addition, allowed and preferred for practical purposes, one may exploit bright squeezed states that can carry high optical powers~\cite{lawrie2019}. In this case, the unit quantum enhancement per photon is still smaller when compared to single photons, but the overall quantum enhancement would be much higher due to the use of the increased optical power (i.e., much larger~$N$). Using bright squeezed states would provide practical quantum polarimetry in the high intensity regime, which we leave for future study. Quantum theory for the characterization of an arbitrary polarization in terms of Stokes parameters has recently been discussed and would be useful for further development of various quantum polarimetric schemes~\cite{ling2006, goldberg2018, goldberg2019}.

\section*{Acknowledgments}
This study was supported by the Basic Science Research Program through the National Research Foundation (NRF) of Korea and funded by the Ministry of Science and ICT  (Grants No. 2019R1F1A1060582, 2019M3E4A1079666) and Institute of Information \& communications Technology Planning \& Evaluation (IITP) grant funded by the Korea government (MSIT) (No. 2019-0-00296).

\appendix
\section*{Appendix}
\section{Fisher information}\label{apped:FI}
\setcounter{equation}{0}
\renewcommand{\theequation}{A\arabic{equation}}
\setcounter{figure}{0}
\renewcommand{\thefigure}{A\arabic{figure}}
\setcounter{table}{0}
\renewcommand{\thetable}{A\arabic{table}}

\setcounter{table}{0}
\renewcommand{\thetable}{B\arabic{table}}
\begin{table*}[t]
\setlength{\tabcolsep}{5pt} 
\renewcommand{\arraystretch}{1.3} 
\begin{tabular}{ |c|| c|c|c|c|c|}
 \hline
 Source of uncertainty&Value&Unit&Divisor (distribution)&Sensitivity&Standard uncertainty [g~ml$^{-1}$]\\
 \hline
  \hline
 \multicolumn{6}{|c|}{\textbf{[type A]}} \\
 \hline
~$\theta_\text{out}$ obtained by~$f_\text{single}$&0.2708&degree ($^\circ$)&1 (N)&0.293&0.0793\\
 \hline
~$\theta_\text{out}$ obtained by~$f_\text{diff}$&0.1841&degree ($^\circ$)&1 (N)&0.293&0.0539\\
 \hline
~$\theta_\text{out}$ obtained by~$f_\text{DSR}$&0.1763&degree ($^\circ$)&1 (N)&0.293&0.0517\\
 \hline
 \multicolumn{6}{|c|}{\textbf{[type B]}} \\
 \hline
Specific rotation [$\alpha(\lambda)$]&0.41&$\text{ }^\circ \text{dm}^\text{-1} \text{g}^\text{-1} \text{ml}\text{ }$&1 (N)&0.05&0.0205\\
 \hline
Length of cuvette $l$ &$\text{ }0.0005\text{ }$&$\text{dm}$&1.732 (R)&17.07&0.0049\\
 \hline
Input polarization $\theta_\text{in}$&0.03&degree ($^\circ$)&1 (N)&0.586&0.0176\\
 \hline
\end{tabular}
\caption{Uncertainty budgets of type A and B in our study. N and R represent the normal and rectangular distributions, respectively. This table is considered at~$\theta_\text{in}=40^\circ$ and~$C$=0.5 g~ml$^{-1}$.}
\label{tab:uncer_bud_table}
\end{table*}

Here we provide the calculation of the FIs when the photon-number-counting measurement is performed at single-mode H or V, or at both modes, respectively. The FI can be written in terms of the probabilities of the measurement outcomes as
\begin{align}
F_{N_\text{H},N_\text{V}}(\theta_\text{out})&=\sum_{N_\text{H},N_\text{V}}\frac{1}{p(N_\text{H},N_\text{V}\vert \theta_\text{out})}\left(\frac{\partial p(N_\text{H},N_\text{V}\vert \theta_\text{out})}{\partial \theta_\text{out}}\right)^2 \label{eq:FI_two}
\end{align}
when~$\theta_\text{out}$ is estimated from measurement at the two modes. Similarly, when measurement is performed at either mode H or mode V, the FI can be written as
\begin{align}
F_{N_\text{s}}(\theta_\text{out})&=\sum_{N_\text{s}}\frac{1}{p(N_\text{s} \vert \theta_\text{out})}\left(\frac{\partial p(N_\text{s} \vert \theta_\text{out})}{\partial \theta_\text{out}}\right)^2,\label{eq:FI_single}
\end{align}
where~$s\in \{\text{H}, \text{V}\}$. It can be easily shown that the associated probabilities are written by
\begin{align}
p^{\text{(q)}}(N_\text{H},N_\text{V})&= \binom{N}{N_\text{H}}\binom{N-N_\text{H}}{N_\text{V}} (\eta_\text{H}T_{\theta_\text{out}})^{N_\text{H}}(\eta_\text{V}R_{\theta_\text{out}})^{N_\text{V}} \nonumber\\ 
&\quad\quad\quad \times (1-\eta_\text{H}T_{\theta_\text{out}}-\eta_\text{V}R_{\theta_\text{out}})^{N-N_\text{H}-N_\text{V}},\\
p^{\text{(q)}}(N_\text{H})&=\binom{N}{N_\text{H}} (\eta_H T_{\theta_\text{out}})^{N_\text{H}}(1-\eta_\text{H} T_{\theta_\text{out}})^{N-N_\text{H}},\\
p^{\text{(q)}}(N_\text{V})&=\binom{N}{N_\text{V}} (\eta_V R_{\theta_\text{out}})^{N_\text{V}}(1-\eta_\text{V} R_{\theta_\text{out}})^{N-N_\text{V}}
\end{align}
for an~$N$-photon number state of light, and 
\begin{align}
p^{\text{(c)}}(N_\text{H},N_\text{V}) &= e^{-\eta_\text{H} T_{\theta_\text{out}} N}\frac{(\eta_\text{H} T_{\theta_\text{out}} N)^{N_\text{H}}}{N_\text{H}!}
e^{-\eta_\text{V} R_{\theta_\text{out}} N}\frac{(\eta_\text{V} R_{\theta_\text{out}} N)^{N_\text{V}}}{N_\text{V}!},\\
p^{\text{(c)}}(N_\text{H})&=e^{-\eta_\text{H} T_{\theta_\text{out}} N}\frac{(\eta_\text{H} T_{\theta_\text{out}} N)^{N_\text{H}}}{N_\text{H}!},\\
p^{\text{(c)}}(N_\text{V})&=e^{-\eta_\text{V} R_{\theta_\text{out}} N}\frac{(\eta_\text{V} R_{\theta_\text{out}} N)^{N_\text{V}}}{N_\text{V}!}
\end{align}
for a coherent state of light with an average photon number~$N$. Substituting these probabilities into Eqs.~\eqref{eq:FI_two} and \eqref{eq:FI_single}, one can write the FIs for the quantum scheme as 
\begin{align}
F_{N_\text{H},N_\text{V}}^{\text{(q)}}(\theta_\text{out}) &= N\left(\frac{\eta_\text{H}}{T_{\theta_\text{out}}} +\frac{\eta_\text{V}}{R_{\theta_\text{out}}}\right)\left(\frac{\partial T_{\theta_\text{out}}}{\partial \theta_\text{out}}\right)^2 \nonumber\\
&\quad\quad+\frac{N(\eta_\text{H}-\eta_\text{V})^2}{1-\eta_\text{H}T_{\theta_\text{out}}-\eta_\text{V}R_{\theta_\text{out}}}\left(\frac{\partial T_{\theta_\text{out}}}{\partial \theta_\text{out}}\right)^2,\label{eq:FI_c_two}\\
F_{N_\text{H}}^{\text{(q)}}(\theta_\text{out}) &= \frac{N\eta_\text{H}}{T_{\theta_\text{out}}(1-\eta_\text{H}T_{\theta_\text{out}})}\left(\frac{\partial T_{\theta_\text{out}}}{\partial \theta_\text{out}}\right)^2,\\
F_{N_\text{V}}^{\text{(q)}}(\theta_\text{out}) &= \frac{N\eta_\text{V}}{R_{\theta_\text{out}}(1 - \eta_\text{V}R_{\theta_\text{out}})}\left(\frac{\partial T_{\theta_\text{out}}}{\partial \theta_\text{out}}\right)^2,
\end{align}
while for the classical scheme as
\begin{align}
F_{N_\text{H},N_\text{V}}^{\text{(c)}}(\theta_\text{out}) &= N\left(\frac{\eta_\text{H}}{T_{\theta_\text{out}}}+\frac{\eta_\text{V}}{R_{\theta_\text{out}}}\right)\left(\frac{\partial T_{\theta_\text{out}}}{\partial \theta_\text{out}}\right)^2,\\
F_{N_\text{H}}^{\text{(c)}}(\theta_\text{out}) &= \frac{N\eta_\text{H}}{T_{\theta_\text{out}}}\left(\frac{\partial T_{\theta_\text{out}}}{\partial \theta_\text{out}}\right)^2,\\
F_{N_\text{V}}^{\text{(c)}}(\theta_\text{out}) &= \frac{N\eta_\text{V}}{R_{\theta_\text{out}}}\left(\frac{\partial T_{\theta_\text{out}}}{\partial \theta_\text{out}}\right)^2,\label{eq:FI_q_Hv}
\end{align}
where~$(\partial_{\theta_\text{out}} T_{\theta_\text{out}})^2 =4 T_{\theta_\text{out}}R_{\theta_\text{out}}~$ and~$F(\theta_\text{out})=F(T_{\theta_\text{out}})\times (\partial_{\theta_\text{out}} T_{\theta_\text{out}})^2$ is used with~$F(T_{\theta_\text{out}})$ being the FI for the parameter~$T_{\theta_\text{out}}$ for mathematical convenience. It is clear that the FI for the quantum scheme with a single-mode measurement is always greater than that for the classical scheme regardless of the values of~$T_{\theta_\text{out}}$,~$R_{\theta_\text{out}}$, and~$\eta_\text{H,V}$. The quantum scheme with two-mode measurement, on the other hand, is beneficial as compared to the classical scheme only when~$\eta_\text{H}\neq\eta_\text{V}$. Note again that the FIs are the same as the QFIs for both classical and quantum schemes considered in this work, so the lower bounds to the uncertainty associated with Eqs.~\eqref{eq:FI_c_two}-\eqref{eq:FI_q_Hv} are ultimately optimal within the framework of the CR inequality.

\section{Uncertainty budget}\label{uncer_bud}
\setcounter{equation}{0}
\renewcommand{\theequation}{B\arabic{equation}}
\setcounter{figure}{0}
\renewcommand{\thefigure}{B\arabic{figure}}

Here we provide the uncertainty budget for type A and B uncertainties~\cite{jcgm2008}, which have the potential to influence the estimation of concentration~$C$. The relevant quantities in this work are the wavelength-dependent specific rotation, the length of the cuvette, and the angles such as~$\theta_\text{in}$ and~$\theta_\text{out}$.
For the specific rotation, we have used a fixed value of the wavelength in Eq.~\eqref{eq:spec_rot}, but in reality no light is entirely monochromatic (i.e., see Fig.~\ref{fig:setup_spectrum}(b)), so that modulation of the wavelength in the reference calculation of the specific rotation needs to be considered in the estimation. Also, any realistic measurement of the length of the cuvette is far away from ideal, so the length might be slightly different from the assumed value of 0.1~dm. A similar behavior would be observed in the calibration of~$\theta_\text{in}$. These effects are treated as type B uncertainty, whereas the measurement of~$\theta_\text{out}$ is actually repeated, leading to type A uncertainty evaluation. Such relevant uncertainties are listed in Table~\ref{tab:uncer_bud_table} for further information.

\section{Monte-Carlo simulation}\label{MCsim}
\setcounter{equation}{0}
\renewcommand{\theequation}{C\arabic{equation}}
\setcounter{figure}{0}
\renewcommand{\thefigure}{C\arabic{figure}}
To understand the effect of experimental imperfections involved in the results shown in Figs.~\ref{fig:preci_enhance} and \ref{fig:concent}, we perform the Monte-Carlo simulation with varying the relevant parameters in an arbitrary range, which are typically constrained in a real experiment. The concentration is estimated from the data obtained by the numerical simulation with varying the input polarization~$\theta_\text{in}$ from~$0^\circ$ to~$10^\circ$ in steps of 1$^\circ$, providing a detailed study in the shaded area in Figs.~\ref{fig:preci_enhance} and \ref{fig:concent}. In order to see the effect of the PE in the estimation, we compare the case of an ideal extinction ratio ($r_\text{PE}\rightarrow\infty$) with the case of a realistic extinction ratio ($r_\text{PE}=1000/1$) for~$C$=0.1, 0.3, and 0.5 g~ml$^{-1}$, shown in Fig.~\ref{fig:MC_sim_fig}. 
It is shown that the estimated concentration is equal to the true value up to the numerical precision of sampling in the ideal case, whereas it drops down at~$\theta_\text{out}=\theta_\text{in}+\alpha\approx0^\circ$, the effect of PE is maximal, indicating an erroneous estimation in the realistic case. The effect of PE becomes less significant as~$T_{\theta_\text{out}}$ approaches 1/2 since the PE of H- and V-photons takes place almost symmetrically and thus compensates each other. Furthermore, note that a little deviation from the true value occurs at~$\theta_\text{out} \approx 0^{\circ}$ even in the ideal case. Such an error can be shown to be independent of~$\nu$ and~$\mu$, but rather to be a technical problem related to using the inverse function of Eq.~\eqref{eq:f_DSR} to estimate~$\theta_\text{out}$. It has been overwhelmed by the effect of PE ({$r_\text{PE}\sim$1000/1) in the measured data shown in Figs.~\ref{fig:preci_enhance} and \ref{fig:concent}.
\begin{figure}[t]
\centering
\includegraphics[width=0.5\textwidth]{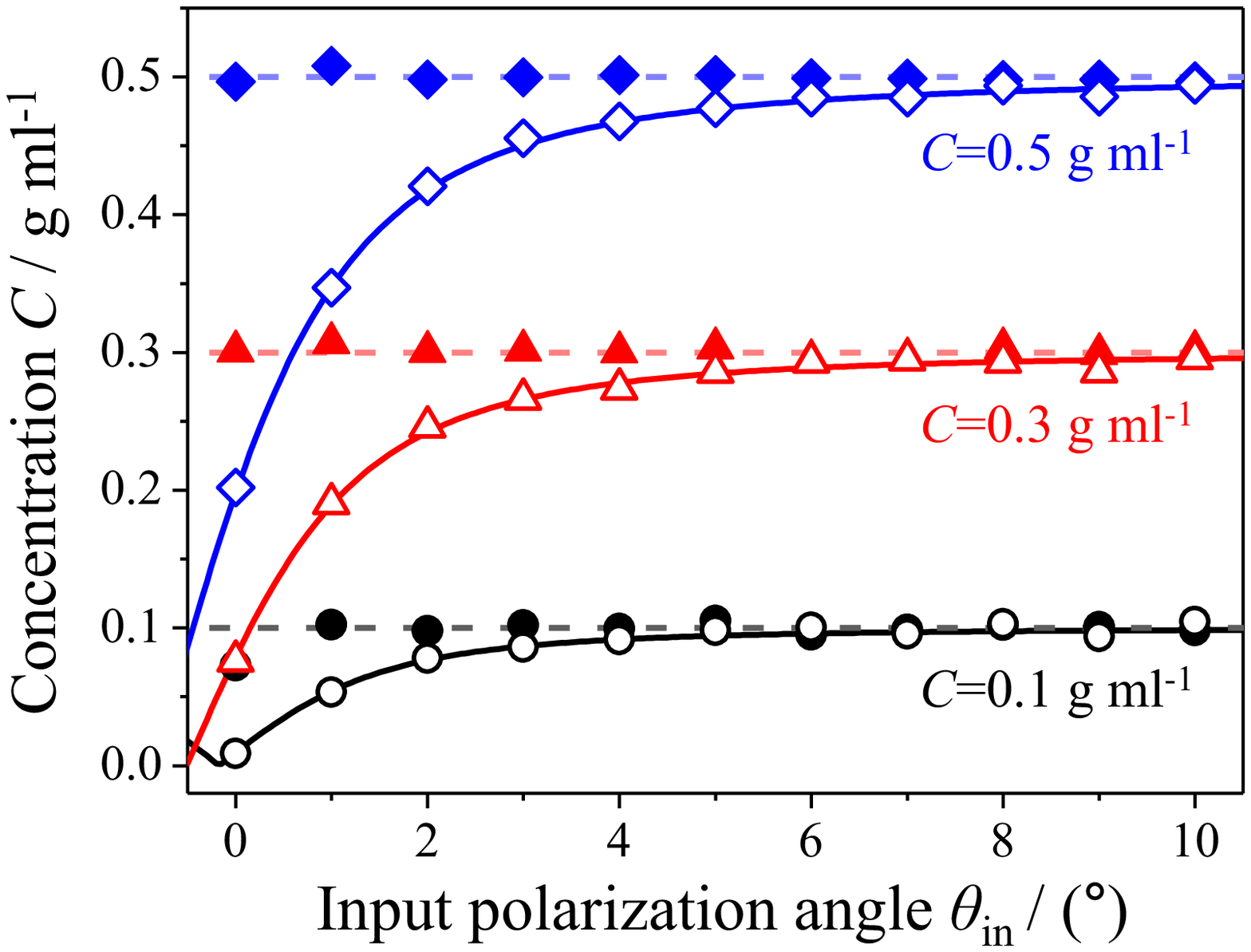}
\caption{
Results of Monte-Carlo simulation to estimate the concentration~$C$ using Eq.~\eqref{eq:f_DSR} for two cases: $r_\text{PE}=1000/1$ (open symbols) and~$r_\text{PE}\rightarrow\infty$ (closed symbols). Solid lines represents the analytical value obtained using Eqs.~\eqref{eq:Rev_f_DSR}, \eqref{eq:delta_f_DSR} and \eqref{eq:delta_theta}, whereas dashed lines are true values of the concentration ($C = 0.1, 0.3$, and $0.5$~g~ml$^{-1}$) used in the simulation. Here,~$\eta_\text{H,V}=0.25$,~$\nu=10^5$, and~$\mu=500$ are used.
}
\label{fig:MC_sim_fig}
\end{figure}

The observed behaviors shown in Fig.~\ref{fig:MC_sim_fig} can be explained via an appropriate modification to the theoretical development made in the main text. The effect of PE, i.e., the erroneous exchange between the H-photon and the V-photon with a rate $r_\text{PE}$, can be included in theory by replacing~$T_{\theta_\text{out}}$ by~$T_{\theta_\text{out}}+r_\text{PE}\times\delta T_{\theta_\text{out}}$, where~$\delta T_{\theta_\text{out}}=1-2T_{\theta_\text{out}}$.  Equations.~\eqref{eq:f_single}-\eqref{eq:f_DSR} can thus be written up to the first order of~$r_\text{PE}\ll1$ as
\begin{align}
\text{(i)\quad\quad\quad}\text{  }\frac{\ave{N_\text{H}}}{\nu}&\approx\left\langle{f_\text{single}}\right\rangle +r_\text{PE} \times\delta f_\text{single}, \label{eq:Rev_f_single}\\
\text{(ii)  }\text{ }\text{ }\frac{\ave{N_\text{H}-N_\text{V}}}{\nu}&\approx\left\langle{f_\text{diff}}\right\rangle+r_\text{PE} \times\delta f_\text{diff},\label{eq:Rev_f_diff}\\
\text{(iii) }\text{ }\ave{\frac{N_\text{H}-N_\text{V}}{N_\text{H}+N_\text{V}}}
&\approx \left\langle{f_\text{DSR}}\right\rangle + r_\text{PE} \times\delta f_\text{diff},\label{eq:Rev_f_DSR}
\end{align}
where
\begin{align}
\delta f_\text{single} &= (1-2T_{\theta_\text{out}})\eta_\text{H},\label{eq:delta_f_single}\\
\delta f_\text{diff} &=(1-2T_{\theta_\text{out}}) (\eta_\text{H}+\eta_\text{V}),\label{eq:delta_f_diff} \\
\delta f_\text{DSR} &=(1-2T_{\theta_\text{out}}) \frac{2 \eta_\text{H}\eta_\text{V}}{[T_{\theta_\text{out}} \eta_\text{H}+(1-T_{\theta_\text{out}}) \eta_\text{V}]^2}.\label{eq:delta_f_DSR}
\end{align}
It is clear to see that the effect of the PE for the three estimators is maximal when~$T_{\theta_\text{out}}=0$ or~$1$, whereas it vanishes when $T_{\theta_\text{out}}=1/2$.

All the errors $\delta f$'s in Eqs.~\eqref{eq:delta_f_single}-\eqref{eq:delta_f_DSR} lead to an equal amount of the error in the estimation of~$\theta_\text{out}$, i.e., $\theta_\text{out}\rightarrow\theta_\text{out}+r_\text{PE}\delta \theta_\text{out}$, where the error 
\begin{align}
\delta \theta_\text{out}=\frac{\cos 2\theta_\text{out}}{\vert \sin 2\theta_\text{out}\vert}=\frac{2T_{\theta_\text{out}}-1}{2\sqrt{T_{\theta_\text{out}}(1-T_{\theta_\text{out}})}}.
\label{eq:delta_theta}
\end{align}
The error $\delta \theta_\text{out}$ caused by the PE diverges as $T\rightarrow 0$ or $1$, while decreases as $T_{\theta_\text{out}}$ approaches $1/2$.

The PE consequently causes an error in the uncertainty $\Delta\theta_\text{out}$ of Eqs.~\eqref{eq:pre_sing_q} - \eqref{eq:pre_DSR_q}, i.e., $\Delta\theta_\text{out}\rightarrow\Delta\theta_\text{out}+r_\text{PE}\times\delta\Delta\theta_\text{out}$, where
$\delta\Delta\theta_\text{out}$ can be found as
\begin{align}
\delta\Delta\theta_\text{out}^\text{single(q)}&=
\frac{1}{\sqrt{\nu}}\frac{1-2T_{\theta_\text{out}}}{8}\left(\frac{1}{\eta_\text{H}}-1\right)
\frac{1}{(1-T_{\theta_\text{out}})^2}
\frac{1}{\Delta\theta_\text{out}^\text{single(q)}},\\
\delta\Delta\theta_\text{out}^\text{diff(q)}&=
\frac{1}{\sqrt{\nu}}\frac{1-2T_{\theta_\text{out}}}{8}
\frac{1}{(\eta_\text{H}+\eta_\text{V})^2}\nonumber\\
&\quad\times\left(\frac{\eta_\text{H}(1-\eta_\text{H})}{(1-T_{\theta_\text{out}})^2}-\frac{\eta_\text{V}(1-\eta_\text{V})}{T_{\theta_\text{out}}^2}\right)
\frac{1}{\Delta\theta_\text{out}^\text{diff(q)}},\\
\delta\Delta\theta_\text{out}^\text{DSR(q)}&=
\frac{1}{\sqrt{\nu}}\frac{1-2T_{\theta_\text{out}}}{8}
\left(\frac{1}{\eta_\text{V}}-\frac{1}{\eta_\text{H}}\right)\frac{1}{\Delta\theta_\text{out}^\text{DSR(q)}}.
\end{align}

\bibliography{referenceabbr.bib}
\end{document}